\documentclass[conference]{IEEEtran}

\usepackage{amsmath,amsfonts,amssymb,varioref,mathrsfs,verbatim}
\usepackage{graphics}
\usepackage{graphicx}
\usepackage{epsfig}
\usepackage{subcaption}
\usepackage[font=small,labelfont=bf, figurename=Fig.]{caption} 
\usepackage{color}
\begin{document}

\title{ Relay-Aided Channel Estimation for mmWave Systems with Imperfect Antenna Arrays}

\author{\IEEEauthorblockN{Mohammed E. Eltayeb}
\IEEEauthorblockA{Department of Electrical and Electronic Engineering\\
California State University, Sacramento\\
Email: mohammed.eltayeb@csus.edu}
}


\maketitle

\begin{abstract}
Compressed Sensing (CS) based channel estimation techniques  have recently emerged as
an effective way to  acquire the channel of  millimeter-wave (mmWave) systems with a small number of measurements. These techniques, however, are based on prior knowledge of transmit and receive array manifolds, and assume perfect antenna arrays at both the transmitter and the receiver. In the presence of antenna imperfections, the geometry and response of the arrays are modified. This distorts the CS measurement matrix and results in channel estimation errors.  This paper studies the effects of both transmit and receive antenna imperfections on the mmWave channel estimate. A relay-aided solution which corrects for errors caused by faulty transmit arrays is then proposed. Simulation results demonstrate the effectiveness of the proposed solution and show that comparable channel estimates can be obtained when compared to systems with perfect antennas without the need for additional training overhead.




\end{abstract}



\section{Introduction}
Communication over the millimeter-wave (mmWave) band is one promising solution to overcome the spectrum crunch and support next generation wireless systems \cite{m1}-\cite{m2}. To provide sufficient link budget for these systems, large antenna arrays need to be deployed at both the base station (BS) and mobile station (MS) \cite{tr}.  The design of the precoding and combining matrices for these systems is contingent on the mmWave channel state information, which is challenging to obtain. This is mainly due to hardware constraints, high-path loss, and antenna faults \cite{ao}-\cite{mg2}. Moreover, antenna faults caused by, for example, blockages (partial or complete), synchronization errors, and/or simply antenna element failures, randomize the array geometry and result in uncertainties in the mmWave channel. These challenges motivate the design of efficient mmWave channel estimation techniques which are independent to changes in the array manifold.

There are mainly two approaches undertaken for channel estimation, (i) beam training \cite{bf1}-\cite{bf7} and (ii) compressed channel estimation (CS) \cite{ao}, \cite{mcs1}-\cite{mcs5}, with the latter favored due to its low training overhead. Despite the large body of work on this topic, prior work assumes fixed and known array structures at both the  the BS and MS. In practice, antenna imperfections can result in random or time-varying array manifolds \cite{mg2}. These imperfections (or faults) can be caused by weather and atmospheric effects which could land debris on outdoor mmWave antennas. Moreover, random finger placement on handheld devices could also lead to random antenna blockages. Faults due to partial or complete antenna blockages randomize the array manifold and introduce errors in the mmWave channel estimate. 

In this paper, we (i) investigate the effects of antenna faults on the mmWave channel estimate, (ii) introduce a relay-aided and CS based channel estimation technique for mmWave systems, and (iii) propose an antenna array diagnosis algorithm.  The introduced CS channel estimation technique  accounts for the mmWave hardware constraints and the randomized array manifold. Additionally, unlike \cite{ao} and \cite{mcs4}, the proposed formulation permits the decoupling of the precoder and combiner design during the channel estimation phase. This is important as it allows the BS to broadcast independent training pilots (or beams) throughout the channel estimation process.  The antenna diagnosis algorithm proposed in this work is different from \cite{mg1}, and \cite{mg2}, as diagnosis is performed at a remote receiver using far-field measurements, and not at the BS (location of faulty antenna) as done in \cite{mg1}, and \cite{mg2}.  Moreover, the proposed algorithm exploits channel training pilots for array diagnosis, and hence, does not require additional measurements to be made.

The remaining of this paper is organized as follows. In Section II, we present the system model. In Section III, we formulate the mmWave channel estimation problem  and demonstrate the effects of antenna imperfections on the compressed mmWave channel estimate. A relay-aided solution to mitigate the effects of BS antenna imperfections is then proposed in Section IV. In Section V we provide some numerical results and conclude our work in Section VI.

\emph{Notation}: Bold uppercase $\mathbf{A}$ is used to denote matrices, bold lower case $\mathbf{a}$ is used to denote column vectors and non-bold lower case $a$ is used to denote scalar values. The $\ell$th norm, the conjugate transpose, and the transpose of a matrix $\mathbf{A}$  is denoted by $||\mathbf{A}||_\ell$, $\mathbf{A}^*$, and $\mathbf{A}^\mathrm{T}$, respectively. The Kronecker product of $\mathbf{A}$ and $\mathbf{B}$ is denoted by $\mathbf{A}\otimes \mathbf{B}$.  $[\mathbf{A}]_{i,j}$ and $[\mathbf{a}]_{i}$ represent the $i,j$th element of the matrix  $\mathbf{A}$ and  the $i$th element of the vector $\mathbf{a}$.

\section{System Model}
We consider a BS with $N_{\text{BS}}$ antennas and $N_{\text{RF}}$ RF chains communicating with a single MS with $N_{\text{MS}}$ antennas and $N_\text{RF}$ RF chains. Both the BS and the MS communicate via $N_\text{S}$ data streams such that, $N_\text{S}\leq N_\text{RF} \leq N_\text{BS}$ and $N_\text{S}\leq N_\text{RF} \leq N_\text{MS}$ \cite{ao}. On the downlink, the BS applies an $N_\text{RF} \times N_\text{S}$ digital precoder $\mathbf{F}_\text{BB}$ followed by an $N_\text{BS} \times N_\text{RF}$ analog preocder, $\mathbf{F}_\text{RF}$. The sampled transmitted symbol therefore becomes $\mathbf{x} = \mathbf{F}_\text{RF}\mathbf{F}_\text{BB} \mathbf{s}$, where $\mathbf{s} = [s_1, s_2, ..., s_\text{N$_\text{RF}$}]^\mathrm{T}$ is the normalized $N_\text{RF}\times 1$ vector of transmitted symbols.  At the MS, the received signals on all antennas are combined to obtain 
\begin{eqnarray}\label{hr1}
  \mathbf{r} =   \mathbf{W}^*_\text{BB}\mathbf{W}^*_\text{RF}\mathbf{H}\mathbf{x} + \mathbf{W}^*_\text{BB}\mathbf{W}^*_\text{RF} \mathbf{n},
 \end{eqnarray}
where  $\mathbf{W}_\text{RF}$ is an $N_{\text{MS}} \times N_{\text{RF}}$ RF combing matrix,  $\mathbf{W}_\text{BB}$  is an $N_{\text{RF}} \times N_{\text{S}}$ digital combing matrix, $\mathbf{H}$ is the $N_\text{MS} \times N_\text{BS}$ matrix that represents the mmWave channel between the BS and MS, and $\mathbf{n} \sim \mathcal{N}(\mathbf{0},\sigma^2\boldsymbol{\text{I}})$.
A geometric channel model with $L$ scatterers is adopted in this paper \cite{ao}, \cite{rap}, and \cite{rap1}.  Under this model, the channel can be expressed as
	\begin{eqnarray}\label{channelk}
	\mathbf{H} = \sqrt{\frac{N_{\text{BS}}N_{\text{MS}}}  {L}} \sum_{\ell=1}^L \alpha_{\ell} {\mathbf{a}}_\text{MS}(\theta_{\ell}){\mathbf{a}}_\text{BS}^*(\phi_{\ell}),
	\end{eqnarray}
where $\alpha_{\ell} \sim \mathcal{CN} (0,1)$ is the complex gain of the $\ell$th path, $\phi_{\ell} \in [0, 2\pi]$ and $\theta_{\ell} \in [0, 2\pi]$ are the $\ell$th path's azimuth angles of departure or arrivals (AoD/AoA) of the BS and the MS. The vector  ${\mathbf{a}}_\text{BS}(\phi_{\ell})$  represents the BS array response while the vector $  {\mathbf{a}}_\text{MS}(\theta_{\ell})$  represents the MS array response. The BS and MS are assumed to know the fault-free geometry of their antenna arrays. While the proposed formulation can be generalized to arbitrary antenna architectures, for ease of exposition, uniform linear arrays (ULAs) with a single RF chain will be assumed throughout this paper.

\section{Formulation of the mmWave Channel Estimation Problem}
In this section we first formulate the proposed channel estimation technique (at the MS) assuming perfect BS and MS antennas. In the second section, we take a more practical approach and highlight the effects of antenna imperfections on the mmWave channel estimate.


\subsection{Channel Estimation with Perfect Transmit and Receive Antennas}
To initiate the channel estimation process, the BS broadcasts training symbols using $M_\text{BS}$ random beams in successive $M_\text{BS}$  time instants.  The MS forms $M_\text{MS}$ random beams, and uses a single beam (or antenna weights) to combine $M=M_\text{BS}/M_\text{MS}$ received training symbols. In the case of perfect transmit and receive antenna arrays, the received signal at the MS receiver becomes
\begin{eqnarray}\label{u01}
{y}_{m,n}=  \mathbf{q}_m^*\mathbf{H}\mathbf{p}_n {s_\text{t}}_n   +  \mathbf{q}_m^*\mathbf{e}_{m,n},
\end{eqnarray} 
where,  $\mathbf{q}_m\in \mathcal{C}^{ N_\text{MS}\times 1}$ is the combining vector (antenna weights) at the MS,  $\mathbf{p}_n \in \mathcal{C}^{N_\text{BS} \times 1}$ is the BS beamforming vector,  ${s_{\text{t}n}}=1$ is the training symbol on the beamforming vector $\mathbf{p}_n$, and $\mathbf{e}\sim\mathcal{N} (0,\sigma^2 \mathbf{I})$ is the $N_\text{MS} \times 1$ noise vector.

Let $\mathbf{Q}=[\mathbf{q}_1, \mathbf{q}_2, ..., \mathbf{q}_{M_\text{MS}}]$ be the $N_\text{MS} \times M_\text{MS}$ measurement matrix at the MS, and  $\mathbf{P}_m=[\mathbf{p}_1, \mathbf{p}_2, ..., \mathbf{p}_{M_\text{}}]$ be the $m$th BS $N_\text{BS} \times M$ beamforming matrix. After $M$ time instances, the received vector at the MS can be written as
\begin{eqnarray}\label{u01v}
  \mathbf{y}_m &=&  (\mathbf{q}_m^* \mathbf{H} \mathbf{P}_m   + \mathbf{q}_m^*\mathbf{E}_m)^\mathrm{T}\\
  &=& (\mathbf{q}_m^* \otimes \mathbf{P}_m^\mathrm{T} ) \text{vec}(\mathbf{H}^\mathrm{T})   +  \mathbf{E}_m^\mathrm{T} \mathbf{q}_m^{*^\mathrm{T}},
\end{eqnarray}
where the $m$th received vector $\mathbf{y}_m \in \mathcal{C}^{M\times 1}$, and matrix  $\mathbf{E}_m \in \mathcal{C}^{N_\text{MS}\times M}$ represents the additive noise.
After $M_\text{BS}$ snapshots, the received vector at the MS becomes
\begin{eqnarray}\label{u021}
\underbrace{\begin{bmatrix}
         \mathbf{y}_1   \\
        \mathbf{y}_2 \\
         \vdots \\
      \mathbf{y}_{M_\text{MS}}
     \end{bmatrix}}_{\mathbf{y}_\text{MS}}
     = \underbrace{
          \begin{bmatrix}
        \mathbf{q}_1^* \otimes \mathbf{P}_1^\mathrm{T}   \\
             \mathbf{q}_2^* \otimes \mathbf{P}_2^\mathrm{T}    \\
         \vdots \\
          \mathbf{q}_{M_\text{MS}}^* \otimes \mathbf{P}_{M_\text{MS}}^\mathrm{T} 
     \end{bmatrix} }_\mathbf{\Psi}
     \text{vec}(\mathbf{H}^\mathrm{T})+
    \underbrace{ \begin{bmatrix}
 \mathbf{E}_1^\mathrm{T} \mathbf{q}_1^{*^\mathrm{T}}   \\
\mathbf{E}_2^\mathrm{T} \mathbf{q}_2^{*^\mathrm{T}} \\
         \vdots \\
    \mathbf{E}_{M_\text{MS}}^\mathrm{T} \mathbf{q}_{M_\text{MS}}^{*^\mathrm{T}}  
     \end{bmatrix} 
     }_\mathbf{e}.
\end{eqnarray}
%
%
Comparing the measurement matrix $\mathbf{\Psi}$ in (\ref{u021}) with the measurement matrix in \cite{ao} and \cite{mcs4} ($\mathbf{\Psi}_\text{A} = \mathbf{P}_1^\mathrm{T} \otimes \mathbf{Q}^*$), we note that the total number of  training beams that result from $\mathbf{\Psi}_\text{A}$ is  $M_\text{T}= M \times M_\text{MS} = M_\text{BS}$ and the total number of independent BS training beams is $\frac{M_\text{BS}}{M_\text{MS}}$. In the proposed formulation (\ref{u021}), the total number of independent BS training beams is  $M_\text{T}= M_\text{BS}$. This is particularly important, especially in the  multi-user channel training case, as it allows the BS to continuously broadcast training beams, irrespective of the number of users or the size of their combining matrices. 

Assuming that all AoAs and AoDs are taken from a grid of $G_\text{BS}$ and $G_\text{MS}$ points, respectively,  and neglecting the grid discretization error, we can approximate  $\mathbf{y}_\text{MS}$ in (\ref{u021}) by \cite{ao}
 \begin{eqnarray}\label{u03}
  \mathbf{y}_\text{MS}= \underbrace{\mathbf{\Psi} (\mathbf{A}_\text{BS}^{*^\mathrm{T}}\otimes \mathbf{A}^{\mathrm{T}}_\text{MS})}_{\text{CS measurement matrix}\boldsymbol{\Phi}} \underbrace{\mathbf{z}}_\text{sparse}  +  \underbrace{\mathbf{e}}_\text{noise},
\end{eqnarray}
where the $N_\text{BS} \times G_\text{BS}$ matrix $\mathbf{A}_\text{BS}$ and $ N_\text{MS} \times G_\text{MS}$ matrix  $\mathbf{A}_\text{MS}$  are the dictionary matrices that consist of the column vectors $\mathbf{a}_\text{BS}(\theta_k)$  and $\mathbf{a}_\text{MS}(\phi_l)$. The   vector $\mathbf{z}$  is a $ G_\text{BS} G_\text{MS} \times 1$ sparse vector which carries the path gains of the corresponding quantizied directions. Applying any off-the-shelf CS recovery algorithms, see example \cite{cb1}-\cite{SL}, one can recover the sparse vector $\mathbf{z}$ from $  \mathbf{y}_\text{MS}$ and  $\boldsymbol{\Phi}$ with a few measurements. 

In the presence of array failures, the entries of measurement matrix $\boldsymbol{\Phi}$ will depend on (i) the density of blockages (ii) failure size (number of imperfect antenna elements), and (iii) failure location (i.e. at BS or MS or both). The effect of failures on the channel estimate is undertaken in the following section.

\begin{figure}[t]
\hspace{0mm}\includegraphics[width=3.3in]{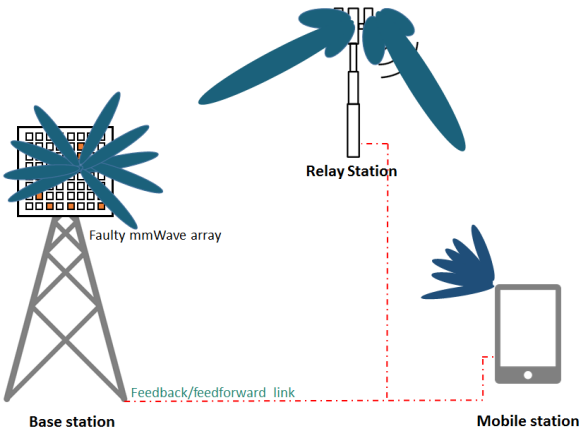}
\caption{An example of an mmWave channel estimation model in which a BS forms random beams to broadcast training pilots. The pilots are exploited by the MS for channel estimation, and by the RS for BS antenna diagnosis. The RS forwards the identity of the faulty antennas and their corresponding weight ($b_i$) to both the BS and the MS via a low rate feedback channel. MS uses diagnostic results to correct for the antenna imperfections and feeds back optimal TX precoder to BS.}
\label{fig1a}
\end{figure}

\subsection{Effects of Transmit and Receive Antenna Imperfections on the Channel Estimate}
In the presence of antenna imperfections (e.g. blockages or failures), the received signal at the MS becomes (see (\ref{u01v}))
\begin{eqnarray}\label{u01vd}
  \mathbf{y}_m &=&  (\mathbf{q}_m^* \mathbf{\hat{H}} \mathbf{P}_m   + \mathbf{q}_i^*\mathbf{\hat{E}}_m)^\mathrm{T},
\end{eqnarray}
where 
\begin{eqnarray}\label{u01vd1}
\hat{\mathbf{H}}  &=&  \sqrt{\frac{N_{\text{BS}}N_{\text{MS}}}  {L}} \sum_{\ell=1}^L \alpha_{\ell} \mathbf{B}_\text{MS}{\mathbf{a}}_\text{MS}(\theta_{\ell})(\mathbf{B}_\text{BS}{\mathbf{a}}_\text{BS}(\phi_{\ell}))^*\\
&=&  \sqrt{\frac{N_{\text{BS}}N_{\text{MS}}}  {L}} \sum_{\ell=1}^L \alpha_{\ell} \mathbf{B}_\text{MS}{\mathbf{a}}_\text{MS}(\theta_{\ell}){\mathbf{a}}^*_\text{BS}(\phi_{\ell}) \mathbf{B}^*_\text{BS}\\
&=& \mathbf{B}_\text{MS}\mathbf{H}\mathbf{B}^*_\text{BS},
\end{eqnarray}
and $\mathbf{\hat{E}}_m = \mathbf{B}_\text{MS} \mathbf{{E}}_m$. The random diagonal matrices $\mathbf{B}_\text{BS}$  and $\mathbf{B}_\text{MS}$ result from imperfections on the BS and MS antenna elements, respectively. The $i$th diagonal entry of the diagonal matrices is defined by
  \begin{equation}\label{efbp1}
b_i  = \left\{
               \begin{array}{ll}
               \alpha_i, & \hbox{ if the $i$th element is blocked}  \\
               1, & \hbox{ otherwise,  } \\
               \end{array}
               \right.
\end{equation}
where  $\alpha_i = \kappa_i e^{j\Phi_i}$,  $0 \leq \kappa_{i} \le 1$ and $0 \leq \Phi_{i} \leq 2\pi$ are the resulting blockage absorption and scattering coefficients at the $n$th element.  A value of $\kappa_{i} = 0$ represents maximum absorption (or blockage) at the $i$th element, and the scattering coefficient $\Phi_{i}$ measures the phase-shift caused by blockages on the $i$th element. Considering the effects of blockages, (\ref{u021}) becomes
\begin{eqnarray}\label{ub2m}
\mathbf{y}_\text{MS}&=& \mathbf{\Psi}  \left(\text{vec}({\hat{\mathbf{H}}^\mathrm{T}})\right) +\mathbf{\hat{e}}\\
&=& \mathbf{\Psi} \left(\text{vec}(\mathbf{B}^{*^\mathrm{T}}_\text{BS}\mathbf{H}^{\mathrm{T}}\mathbf{B}_\text{MS}^\mathrm{T})\right) +\mathbf{\hat{e}}\\ \label{ub2mi}
&=& \mathbf{\Psi}  \left( \mathbf{B}_\text{MS} \otimes \mathbf{B}_\text{BS}^{*^\mathrm{T}}\right) \left(\text{vec}({\mathbf{{H}^\mathrm{T}}})\right)   +{\hat{\mathbf{e}}},
\end{eqnarray}
where $\mathbf{\hat{e}} = [\mathbf{q}_{1}^* \hat{\mathbf{E}}_1,  ...,  \mathbf{q}_{M_\text{MS}}^* \hat{\mathbf{E}}_{M_\text{MS}}]^{\mathrm{T}}$. Neglecting the grid discretization error, (\ref{ub2mi}) can be written as
\begin{eqnarray}\label{u2vm}
  \mathbf{y}_\text{MS}= \underbrace{\mathbf{\Psi}  \left( \mathbf{B}_\text{MS} \otimes \mathbf{B}_\text{BS}^{*^\mathrm{T}}\right) \left(\mathbf{A}_\text{BS}^{*^\mathrm{T}}\otimes \mathbf{A}^{\mathrm{T}}_\text{MS}\right) }_{\text{corrupted CS measurement matrix}\hat{\boldsymbol{\Phi}}} \underbrace{\mathbf{z}}_\text{sparse}  +  \underbrace{\hat{\mathbf{e}}}_\text{noise}.
\end{eqnarray}
Comparing  (\ref{u2vm}) with (\ref{u03}), we observe that random array failures corrupt the CS measurement matrix, or equivalently modify the BS and MS array responses, and introduce channel estimate errors during CS recovery. In the next section, we introduce an array diagnosis technique that is able to detect the locations and corresponding block coefficients $b_i$. This allows the MS to mitigate the effects of antenna imperfections. 


\section{Relay-Aided Channel Estimation}
This section introduces the proposed relay-aided channel estimation technique. Prior to that,  we make the following assumptions: (i) a fixed relay station (RS), with a line-of-sight (LoS) link, aids both the BS and the MS during the channel estimation phase only. (ii) For ease of exposition, faults are assumed to be present at the BS antenna only, i.e. $\mathbf{B}_\text{MS}=\mathbf{I}$.  Nonetheless, the proposed technique can be used to detect antenna faults at all network terminals. (iii) Fault locations (and coefficients) are constant during the channel estimation and data transmission interval. (iv) BS fault-free response is known at the relay.

To initiate the channel estimation process, the BS broadcasts $M_\text{BS}$ training beams in $M_\text{BS}$ time instances. These beams are utilized by the MS to estimate its channel with the BS, and simultaneously by the relay to diagnose the BS transmit antenna as shown in Fig. \ref{fig1a}. The $n$th output at the relay is given by
\begin{eqnarray}\label{r1}
{y}_n&=&{\gamma} (\mathbf{B}_\text{BS}\mathbf{h}_\text{r})^*\mathbf{p}_n +  {\epsilon},
\end{eqnarray} 
where $\gamma$ is the relay's receive antenna gain, $\mathbf{h}_\text{r}= \alpha_\text{r} {\mathbf{a}}_\text{BS}(\phi_{r})$,  is the BS-RS channel, $\alpha_\text{r}$ is the BS-RS channel path loss (assumed to be known by the RS), $\phi_{r}$ is the BS-RS AoD/AoA,  and  $\epsilon \sim \mathcal{CN}(0, \sigma^2_\text{r})$ is the additive noise. The matrix $\mathbf{B}_\text{BS}$ in (\ref{r1}) results from imperfections at the BS antenna. After $M_\text{BS}$ measurements, the received vector at the relay can be written as
\begin{eqnarray}\label{rm}
  \mathbf{y}^*&=&{\gamma \mathbf{P}^*\mathbf{B}_\text{BS}\mathbf{h}_\text{r}}+   \boldsymbol{\epsilon}\\  \label{rem2}
    &=&{ \gamma \mathbf{P}^*\mathbf{h}_\text{r}}+ {\gamma \mathbf{P}^*(\mathbf{B}_\text{BS} -\mathbf{I})\mathbf{h}_\text{r}}+   \boldsymbol{\epsilon},
\end{eqnarray}
where $\mathbf{P}=[\mathbf{p}_1, \mathbf{p}_2, ..., \mathbf{p}_{M_\text{BS}}]$. The first term in (\ref{rem2}) represents the error-free pattern at the relay which results from the training beamforming vector $\mathbf{P}$, and the second term represents the (amplified)  error which results from the faulty BS antenna elements. Note that in the ideal case, $\mathbf{B}_\text{BS}=\mathbf{I}$ and the second term  in (\ref{rem2}) becomes zero. 

Subtracting the ideal error-free response from (\ref{rm}) we obtain
\begin{eqnarray}\label{ivg}
  \mathbf{y}_s= \mathbf{y}^* - {\gamma\mathbf{P}^*\mathbf{h}_\text{r}} = \gamma \mathbf{P}^* \mathbf{g} +   \boldsymbol{\epsilon},
\end{eqnarray}
 Note the innovation vector $\mathbf{g}=(\mathbf{B}_\text{BS} -\mathbf{I}) \mathbf{h}_\text{r}$ in (\ref{ivg}) is sparse, with the non-zero elements representing the locations of the faulty antennas.  To estimate the matrix $\mathbf{B}_\text{BS}$, the relay needs to recover the sparse vector $\mathbf{g}$. As the measurement matrix $ \mathbf{P}$ and the error-free measurements, $y_\text{perfect}={{\gamma} \mathbf{P}^*\mathbf{h}_\text{r}}$ are known at the relay, the relay applies  sparse recovery algorithms, e.g. \cite{cb1}-\cite{SL}, to estimate $\mathbf{g}$ from $ \mathbf{y}_s$. Once the vector $\mathbf{g}$ is estimated, the $i$th diagonal entry of $\mathbf{B}_\text{BS}$ can be calculated as follows $[\mathbf{B}_\text{BS}]_{i,i} = \frac{[\mathbf{g}]_i}{[\mathbf{h}_r]_i} +1.$

To complete the channel estimation step, the relay forwards the non-zero entries of the matrix $\mathbf{B}_\text{BS}$ to the MS. The MS uses $\mathbf{B}_\text{BS}$  to (i) form its CS measurement matrix (see (\ref{u2vm})) and estimate its channel with the BS, (ii) form the optimal precoding/combining vectors and forwards this information to the BS. The design of the precoder/combiners is omitted in this paper for brevity.

\begin{figure}[t]
\includegraphics[width=3.7in]{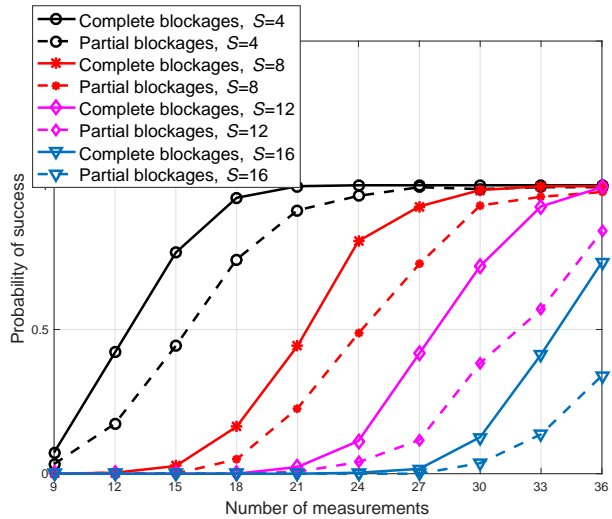}
\caption{ Detection of BS antenna faults at the relay station. BS antenna is subject to complete and partial blockages. For complete blockages ($b_i =0$) and partial blockages $\angle b_i = \frac{\pi}{4}$. Probability of success  increases with the number of measurements. $N_\text{BS}=64$.  }
\label{fig1}
\end{figure}

\section{Simulation Results}
In this section, we conduct numerical simulations to evaluate the performance of the proposed techniques.  We consider a setup where a BS, with a possible faulty antenna, is serving an MS with the aid of a fixed RS with LoS link to the BS. In this setup, the BS, MS and RS are assumed to be equipped with ULAs, each with half wavelength separation. Further, the RS is assumed to have perfect knowledge of the BS error-free array response and  its channel with the BS. The BS-RS link signal-to-noise ratio (SNR) is fixed to 30dB. For the BS-MS link, all AoAs and AoDs are assumed to be quantized and are taken from a grid of $\mathbf{G_\text{BS}}=64$ and $\mathbf{G_\text{MS}}=32$. To generate the random blockages, the values of $\kappa$ and $\Phi$ in (\ref{efbp1}) are chosen uniformly and independently at random from the set  $\{i \in \mathcal{R}: 0 \leq i \leq 1\}$ and $\{0,..,2\pi\}$ respectively.  The entries of the matrices $\mathbf{P}$ and $\mathbf{Q}$ are drawn from $\{\pm1, \pm j \}$ with equal probability. The LASSO \cite{cP08} is implemented for sparse recovery as it does not require the sparsity $S$ to be known a priori. We adopt the success probability  as a performance measure to quantify the error in detecting the faulty antenna element locations, and the  normalized mean square error (NMSE) as a performance measure to quantify the error in  estimating the mmWave channel at the MS. The NMSE is defined by $\text{NMSE} = \frac{  \| \mathbf{H}-\tilde{\mathbf{H}} \|^2_2 }{   \| \mathbf{H} \|^2_2  },$
where $\mathbf{\tilde{H}}$ is the estimated channel at the MS.

To examine the performance of the proposed diagnostic technique the at the relay, we plot the success probability at the RS obtained for several complete and partial blockages in Fig. \ref{fig1}. For all cases, the figure shows, (a) the success probability increases with increasing number of measurements, and (b) higher number of measurements are required to detect faulty antennas with partial blockages. Partial blockages reduce the norm of the innovation vector $\mathbf{g}$ in ($\ref{ivg}$), thus requiring more measurements for recovery. It should be noted that the RS is able to diagnose BS antenna faults, at no extra training overhead, by exploiting the downlink channel training beams (or pilots) intended for channel estimation.


\begin{figure}[t]
\includegraphics[width=3.7in]{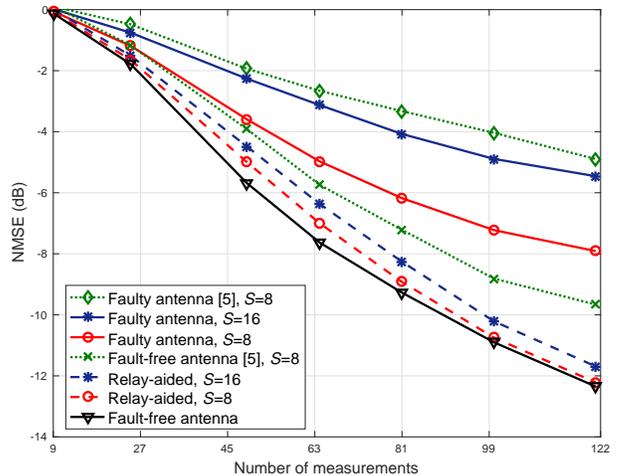}
\caption{Channel estimation at the MS using different estimation techniques. $N_\text{BS}=64$, $N_\text{MS}=32$, $L=3$ paths and the SNR at the MS is 10 dB.}
\label{fig2}
\end{figure}

\begin{figure}[t]
\includegraphics[width=3.7in]{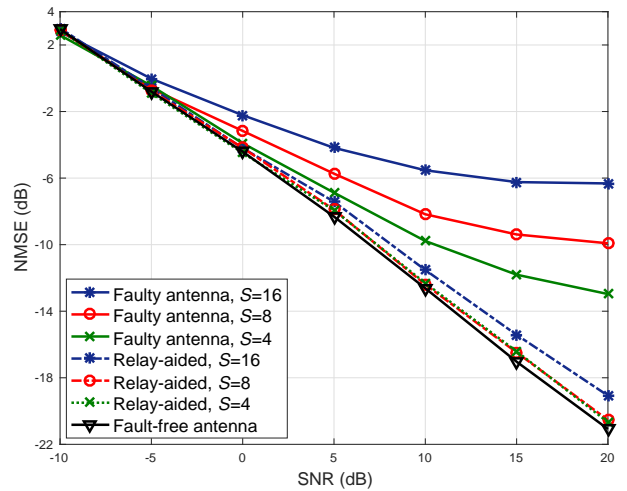}
\caption{Channel estimation at the MS with faulty BS antenna. $N_\text{BS}=64$, $N_\text{MS}=32$, $L=3$ paths and number of measurements is fixed to $M_\text{BS}=121$.}
\label{fig3}
\end{figure}

In Fig. \ref{fig2} the normalized mean squared error of the channel estimates at the MS is shown as a function of the number of compressed sensing measurements.  For comparison, the widely adopted CS channel estimation technique proposed in \cite{ao} (and \cite{mcs4}) is simulated with perfect and imperfect BS antennas. In the presence, and absence, of blockages, Fig. \ref{fig2} shows that the proposed channel estimation formulation (without the aid of the relay) provides a lower NMSE estimate when compared to the formulation proposed in \cite{ao}. This gap stems from the formulation in \cite{ao} which dictates the total number of independent BS training beams to be $\frac{M_\text{BS}}{M_\text{MS}}$. The proposed formulation permits the BS to broadcast $M_\text{BS}$ independent training beams. This enhances the CS measurement matrix and the channel estimate at the receiver.  The impact of the BS antenna faults on the channel estimate is also shown for all cases. For $S=16$ and $S=8$ faulty antenna elements, the figure shows an increase in the NMSE when compared to the fault-free case. Moreover, the NMSE gap does not decrease with increasing number of measurements. Nonetheless, Fig. \ref{fig2} shows that the proposed relay-aided solution provides an NMSE comparable to that obtained by fault-free systems without requiring additional number of measurements. Thanks to the diagnostic results provided by the relay, the MS can perform real-time error-correction on it is CS measurement matrix and use it for sparse channel recovery.

To investigate the effect of the receive SNR on the channel estimate, we plot the NMSE of the proposed technique with varying number of faulty antenna elements in Fig. {\ref{fig3}}. The figure shows the NMSE increases with the number of faults and saturates with increasing SNR. For all cases, the relay-aided solution is shown to provide an NMSE comparable to that obtained by fault-free antennas. The figure also shows that the relay-aided technique experiences a slight performance hit for higher number of blockages. The reason for this is that as the number of blockages increase, more measurements are required at the RS to successfully diagnose the BS antenna (this is evident from Fig. \ref{fig1}). Therefore, for high number of faults, the total number of required measurements becomes a function of the number of antenna faults, in addition to the number of channel paths and quantized AoAs/AoDs, and BS-RS and BS-MS SNR.

\section{Conclusions} \label{sec:con}
In this paper, we investigated the effects of antenna faults on mmWave compressed channel estimate and proposed a new formulation for mmWave channel estimation. We showed that antenna faults modify the antenna geometry and as a result, distort the mmWave channel estimate. To mitigate the effects of faults, we proposed a relay-aided channel estimation technique for these systems. The proposed technique permits far-field diagnostic measurements to be taken from a single location and then broadcasted to the network without the need for any additional training time. When faults exist at the BS, we showed that the proposed technique reliably detects the locations of the faulty antenna elements, if any, and estimates the corresponding attenuation and phase-shift coefficients caused by blockages.  Via simulations, we also showed that the proposed relay-aided solution realizes channel estimates comparable to that obtained by systems with fault-free antennas. 


{}
\end{document}